\pdfoutput=1
\input{aipcheck}

\documentclass[
    ,final            % use final for the camera ready runs
  ]
  {aipproc}

\usepackage{amssymb}
\usepackage{amsmath}
\usepackage{amsfonts}
\usepackage{subfig}
\usepackage{color}

\def\nn{\nonumber}
\newcommand{\df}{\mathrm{d}}

\layoutstyle{8x11single}

\begin{document}

\title{Secondary Production of Massive Quarks in Thrust}


\author{Andr\'e H.~Hoang}{
  address={University of Vienna, Faculty of Physics, Boltzmanngasse 5, A-1090 Wien, Austria},
  altaddress={Erwin Schr\"odinger International Institute for Mathematical
Physics, University of Vienna, Boltzmanngasse 9, A-1090 Vienna, Austria}
}
\author{\underline{Vicent Mateu}}{
  address={University of Vienna, Faculty of Physics, Boltzmanngasse 5, A-1090 Wien, Austria},
  altaddress={Speaker}
}
\author{Piotr Pietrulewicz}{
  address={Theory Group, Deutsches Elektronen-Synchrotron (DESY), D-22607 Hamburg, Germany\\
  [2pt] Preprint numbers\,: UWTHPH 2014-36,~DESY 14-249}
}

\begin{abstract}
We present a factorization framework that takes into account the production of heavy quarks through gluon splitting in
the thrust distribution for $e^+e^-\to$ hadrons. The explicit factorization theorems and some numerical results are displayed 
in the dijet region where the kinematic scales are widely separated, which can be extended systematically to the whole
spectrum. We account for the necessary two-loop matrix elements, threshold corrections, and include resummation up to
N$^3$LL order. We include nonperturbative power corrections through a field theoretical shape function, and remove the
$\mathcal{O}(\Lambda_{\rm QCD})$ renormalon in the partonic soft function by appropriate mass-dependent subtractions.
Our results hold for any value of the quark mass, from an infinitesimally small (merging to the known massless result) to an
infinitely large one (achieving the decoupling limit). This is the first example of an application of a variable flavor
number scheme to final state jets.
\end{abstract}

\maketitle

\section{Introduction}
A systematic and coherent treatment of mass effects in collider observables (such as event shapes) is mandatory to match
the expected accuracy of upcoming experimental data at the LHC and future linear colliders. This encompasses the production
of top quarks and other massive colored particles (yet to be discovered) at very high energies in hadron-hadron
and lepton-lepton machines, bottom quark production at lower energies (e.g.\ LEP, TASSO, etc\ldots), and also charm production 
in lepton-hadron scattering experiments.

Concerning initial state mass effects in inclusive cross sections at hadron colliders, a first description
for arbitrary mass scales has been provided by Aivazis, Collins, Olness and Tung (ACOT) in two seminal
papers~\cite{Aivazis:1993kh,Aivazis:1993pi}. Their results are based on using different renormalization prescriptions
depending on the relation between the quark mass and the hard momentum transfer scale, and lay the foundations for existing
variable flavor number schemes (VFNS). This concept can be incorporated in a convenient way into effective field theories,
as was shown in Ref.~\cite{Pietrulewicz:2014qza,Hoang:2014ira} in the Soft-Collinear Effective Theory (SCET)~\cite{Bauer:2000ew,Bauer:2000yr}
framework.

Based on Refs.~\cite{Gritschacher:2013pha,Pietrulewicz:2014qza} we present here a VFNS applied to
state jets. We address the description of the production of heavy quarks through the secondary radiation off a virtual
gluon. In contrast to the case of primary production of heavy quarks, in which these are created already in the hard 
interaction~\cite{Fleming:2007qr,Fleming:2007xt}, we thus consider here the situation, where massless quarks are
primarily produced in the $e^+e^-$ collision, and one of the subsequent radiated gluons splits into a heavy quark-antiquark
pair before hadronization takes place. The approach we describe here continuously interpolates between the very large mass
limit, where one achieves decoupling of the heavy quark, and the very small mass limit, where one approaches the known
massless quarks results, without upsetting the overall parametric precision of the corresponding massless factorization
theorem. Secondary mass effects are naturally small, since they happen at $\mathcal{O}(\alpha_s^2)$. However they are
conceptually valuable since a number of nontrivial issues concerning the construction of VFNSs are clarified which are also relevant for other applications, where mass effects are more prominent.
The case of secondary heavy quark radiation is also specific since there is no kinematical situation for which the production of the massive quarks needs to be treated in the framework of boosted Heavy Quark Effective Theory (bHQET)~\cite{Pietrulewicz:2014qza}.

We will deal with the thrust event-shape distribution in $e^+e^-$ collisions as a specific example,
emphasizing that the factorization setup can be easily adapted to other event shapes, and with some more effort to
the more complicated environment of hadron collisions. We define thrust as:
\begin{equation}\label{eq:thrust-def}
\tau \,=\, 1-T \,=\,
\min_{\vec{n}} \left( 1 -\, \frac{ \sum_i|\vec{n} \cdot \vec{p}_i|}{\sum_j E_j} \right) =
\min_{\vec{n}} \left( 1 -\, \frac{ \sum_i|\vec{n} \cdot \vec{p}_i|}{Q} \right),
\end{equation}
where $\vec{n}$ is referred to as the thrust axis. Eq.~(\ref{eq:thrust-def}) reduces to the familiar definition of thrust
\cite{Farhi:1977sg} for massless particles which it is not convenient for analytic computations within the
factorization approach described below.
The definition in Eq.~(\ref{eq:thrust-def}) maximizes the effect of heavy particles making it more suitable for a
heavy-quark mass determination~\cite{Fleming:2007qr,Fleming:2007xt}.

The dijet limit is enforced by small values of $\tau$ (which can be thought of a veto on a third jet), and in this
situation the final state constitutes of two narrow back-to-back jets plus soft radiation at large angles. One can
identify three physical scales, with which we associate corresponding renormalization scales:
the hard scale $\mu_H\sim Q$ (scale of the short distance interaction), the jet scale $\mu_J\sim Q\,\lambda$ (typical
transverse momentum of a jet) and the soft scale $\mu_S\sim Q \lambda^2$ (energy of soft particles).
The effective field theory correspondingly has $n$-, ${\bar n}$-collinear and ultra-soft modes with momenta (in light-cone
coordinates) \mbox{$p^\mu_n\sim Q\,(\lambda^2,1,\lambda)$}, \mbox{$p^\mu_{\bar n}\sim Q\,(1,\lambda^2,\lambda)$} and
$p_{\rm us}^\mu \sim Q( \lambda^2,\lambda^2, \lambda^2)$ , respectively. In this limit the dominant part of the cross section 
(which we denote by singular cross section) factorizes into a hard matching coefficient and the convolution of a jet and
soft functions~\cite{Fleming:2007qr,Fleming:2007xt,Schwartz:2007ib,Bauer:2008dt}.
When dealing with secondary production of a pair of quarks with mass $m$ one needs to introduce additional \textit{mass modes},
that have an intrinsic fluctuation scale related to parameter $\lambda_m\equiv m/Q$ associated to their mass $m$,
in addition to the massless power counting parameter $\lambda\sim \tau^{1/2}$. The corresponding $n$-, ${\bar n}$-collinear 
and soft mass modes contain mass-shell fluctuations scaling as \mbox{$p^\mu_n\sim Q\,(\lambda_m^2,1,\lambda_m)$},
\mbox{$p^\mu_{\bar n}\sim Q\,(1,\lambda_m^2,\lambda_m)$} and $p_{\rm s}^\mu \sim Q( \lambda_m,\lambda_m,\lambda_m)$,
respectively. (Note that the soft mass modes have typical momenta scaling linearly with $\lambda_m$.)
In particular the soft and collinear
mass-shell fluctuations have the same virtuality, $p_{\bar n}^2\sim p_{n}^2\sim p_{s}^2\sim m^2$, which leads to the
emergence of rapidity logarithms in the threshold corrections at the mass scale.

\section{Factorization theorem}
Since we are dealing with primary production of massless quarks and secondary production of heavy quarks, we can take the
massless factorization theorem for thrust as our starting point.\footnote{A similar factorization theorem for C-parameter
has been recently derived in Ref.~\cite{Hoang:2014wka} and a general theorem can be found in \cite{Bauer:2008dt}. These
factorization theorems can be easily extended to measure the angle between the thrust axis and the beam 
direction~\cite{Mateu:2013gya}.} It reads~\cite{Korchemsky:1999kt,Schwartz:2007ib,Fleming:2007qr}
\begin{align}
\!\!\!\!\!\frac{1}{\sigma_0}\frac{\df\sigma}{\df\tau}\,=\, Q \,H^{(n_{\!f})}(Q,\mu_H)\,
U_H^{(n_{\!f})}(Q,\mu_H,\mu_S)\!\!
\int\! \df s\!\!\int \! \df s^\prime\, J^{(n_{\!f})}(s^\prime,\mu_J)\,U^{(n_{\!f})}_J(s-s^\prime,\mu_S,\mu_J)
\,\, S^{(n_{\!f})}\!\Big(Q\,\tau-\frac{s}{Q},\mu_S\Big) \, .
\label{eq:diffsigma0}
\end{align}
Here $\sigma_0$ denotes the tree-level $e^+ e^- \to q\bar{q}$ total cross section and $H$, $J$ and $S$ are the hard,
jet, and soft functions, respectively; $U_H$ and $U_J$ are the RG factors for the hard and jet functions. For
definiteness we have chosen to evolve the hard and jet functions to the soft scale (although this choice is arbitrary),
and therefore there is no evolution of the soft function. The dependence on the number of light quark flavors $n_f$
relevant for the RG evolution is explicitly indicated in each of the terms in Eq.~(\ref{eq:diffsigma0}). It starts
already at LL order through the RG evolution factors, which depend on $\alpha_s^{(n_f)}$. The matrix elements receive an explicit dependence on $n_f$ from $\mathcal{O}(\alpha_s^2)$.

The soft function can be further factorized into a partonic soft function $\hat S^{(n_f)}$, calculable in perturbation
theory, and a nonperturbative shape function $F$. This separation into partonic and hadronic components is usually performed in a
scheme relying on dimensional regularization, which leads to an $\mathcal{O}(\Lambda_{\rm QCD})$ renormalon
in $\hat S^{(n_f)}$. This problem can be eliminated introducing a gap parameter and gap subtractions (which depend on
the renormalization scale $\mu_S$ and the gap-subtraction scale $R$) in the definition of the leading power correction
for the OPE region~\cite{Hoang:2007vb,Hoang:2008fs,Mateu:2012nk}. These also depend on the number of massless flavors $n_f$.

The quark mass adds a complication to the factorization setup since it can a priori adopt any hierarchy with respect
to the hard, collinear or soft scales, which themselves are functions of $\tau$. As will be described below, this
leads to renormalization group factors with a variable numbers of active quark flavors. Furthermore, when
crossing the quark mass scale in the evolution, threshold corrections arise that are related to virtual fluctuations
in the hard, collinear and soft sectors as well as to the gap subtractions which are in analogy to the threshold
corrections known in the evolution of $\alpha_s$ and the parton distribution functions. Finally, also mass-dependent fixed-order
corrections in the hard, jet and soft functions have to be taken into account.

\section{Mass Mode setup and different scenarios}
In this section we summarize the mass mode setup of Ref.~\cite{Gritschacher:2013pha}, which is based on four different
scenarios. Each scenario corresponds to a different hierarchy between the quark mass and the hard, collinear and soft
scales. Depending on the relative sizes of $\lambda_m$ and $\lambda$ one of the four scenarios has to be employed. We
emphasize that we do not require large hierarchies of the mass with respect to the other three scales for the
characterization of these scenarios, since neighboring scenarios merge continuously into each other. %Dynamic real
To discuss the various scenarios we consider a generic setup with {\it one} heavy quark with mass $m$ and $n_l$
light quarks. In the following we denote with $\mu_m\sim m$ the scale at which one switches between the schemes which contain either $n_l$ or $n_l+1$ running flavors. The corresponding explicit calculations and results can be found in
Refs.~\cite{Gritschacher:2013pha,Gritschacher:2013tza,Pietrulewicz:2014qza}.
\subsection{Scenario I: $\mathbf{m>Q>Q\,\lambda>Q\,\lambda^{\!2}}$}
If $\mu_m > \mu_H$ the massive quark is integrated out already at the level when SCET is matched onto QCD.
Therefore the massive quark only affects the hard matching
coefficient, but not the jet and soft functions. The factorization theorem is analogous to Eq.~(\ref{eq:diffsigma0}) with
$n_l$ active flavors, except for the hard current matching coefficient which depends on the heavy quark mass through
virtual effects,
\begin{align}
 \frac{1}{\sigma_0} \frac{\df\sigma}{\df\tau}= Q\,H^{(n_l)}(Q,m,\mu_H)\,
U^{(n_l)}_H(Q,\mu_H,\mu_S) \int \!\df s\!  \int\! \df s'\, J^{(n_l)}(s',\mu_J)\, U^{(n_l)}_J(s-s',\mu_S,\mu_J)\,
 S^{(n_l)}\Big(Q\,\tau-\frac{s}{Q},\mu_S\Big).
\label{eq:diffsigmaI}
\end{align}
Here the strong coupling constant in each of the matrix elements or running factors runs with $n_l$ active flavors.
Scenario I shows manifest decoupling in the infinite mass limit:
\begin{align}
 \lim_{m \to \infty} H^{(n_l)}(Q,m,\mu) = H^{(n_l)}(Q,\mu) \, .
\end{align}
Hence in this limit the factorization theorem in Eq.~\eqref{eq:diffsigmaI} reduces to the massless one in
Eq.~(\ref{eq:diffsigma0}) with $n_f = n_l$ active flavors. To achieve this desirable property one needs to renormalize
the massive quark bubble contribution to the QCD form factor with the on-shell (OS) subtraction (i.e. with zero-momentum
subtraction) while the $n_l$ massless quark bubble contributions are still renormalized
in the $\overline{\rm MS}$ subtraction as usual. So the massive quark is not an active running flavor.
The factorization formula Eq.~(\ref{eq:diffsigmaI}) is not appropriate for the limit of small $m$ due to large logarithms
that appear in $H^{(n_l)}(Q,m,\mu)$ for $m\to 0$.
\subsection{Scenario II: $\mathbf{Q>m>Q\,\lambda>Q\,\lambda^{\!2}}$}
In this scenario one has $\mu_H > \mu_m > \mu_J>\mu_S$.
Now the mass modes contribute as dynamic degrees of freedom in SCET above $\mu_m$. For the hard matching coefficient,
one needs to sum up large logs that show up when $Q\gg m$, and make sure that the known massless limit can be smoothly 
attained for $m/Q\to 0$. The former is achieved by evolving $H$ with $n_l + 1$ active flavors from $\mu_H$ to
the mass scale $\mu_m$, where the heavy quark is integrated out, and further evolving down from $\mu_m$ to $\mu_S$ with
$n_l$ active flavors. Both running factors are mass independent. At the matching scale $\mu_m$ the threshold correction 
$\mathcal{M}_H$ has to be included. Mass effects are purely virtual, and their contributions to the jet and soft functions 
vanish identically if the OS scheme for $\alpha_s$ and the matrix elements is used. Therefore the jet function (at the scale 
$\mu_J$) and the soft function (at the scale $\mu_S$) are identical to the massless case with $n_l$ active flavors. The 
factorization theorem reads:
\begin{align}
\frac{1}{\sigma_0}\frac{\df\sigma}{\df\tau} \,=\, & \,Q\, H^{(n_l+1)}(Q,m,\mu_H)\,
U^{(n_l+1)}_{H}(Q,\mu_H,\mu_m) {\mathcal{M}_{H}(Q,m,\mu_m)}\,U^{(n_l)}_{H}(Q,\mu_m,\mu_S) \\
& \times \int\! \df s \! \int \!\df s'\,J^{(n_l)}(s',\mu_J) \, U^{(n_l)}_J(s-s',\mu_S,\mu_J)\, S^{(n_l)}\Big(Q\,\tau-\frac{s}{Q},\mu_S\Big).\nn
\label{eq:diffsigmaII}
\end{align}
Here $H^{(n_l+1)}$ and $U^{(n_l+1)}_{H}$ depend on $\alpha_s^{(n_l + 1)}$ (i.e. in the ($n_l+1$)-flavor scheme), whereas everything else depends on
$\alpha_s^{(n_l)}$ (i.e. in the ($n_l$)-flavor scheme).\footnote{The threshold correction $\mathcal{M}_{H}$ can be displayed in either the ($n_l$)- or the
($n_l+1$)-flavor scheme.} Note that the difference between $H^{(n_l+1)}(Q,m,\mu_H)$ and $H^{(n_l)}(Q,m,\mu_H)$
is not only the different number of active flavors in $\alpha_s$. In the former there are contributions from non-vanishing
SCET diagrams that take part in the matching computation, and also the $\overline{\rm MS}$ scheme has been used.
The massless limit for $H$ occurs naturally in this renormalization scheme,
\begin{equation}
\lim_{m\to 0} H^{(n_l+1)}(Q,m,\mu_H) = H^{(n_l+1)}(Q,\mu_H)\,,
\end{equation}
where the RHS is the hard function appearing in the massless factorization theorem in Eq.~(\ref{eq:diffsigma0}) with
$n_f = n_l+1$. The decoupling limit for the jet and soft function is trivial. The threshold coefficient
$\mathcal{M}_H$ depends on large rapidity logarithms $\log(m^2/Q^2)$ which have to be considered of order $\alpha_s^{-1}$
in the logarithmic counting. They are known to exponentiate and have been computed explicitly up to three
loops~\cite{Ablinger:2014vwa}, which yields necessary ingredients for an overall N$^3$LL resummation.
\subsection{Scenario III: $\mathbf{Q>Q\,\lambda>m>Q\,\lambda^{\!2}}$}
Here one has $\mu_H>\mu_J>\mu_m>\mu_S$\,, and therefore there is no change in the hard matching coefficient, its running
factors and the corresponding threshold coefficient compared to the previous scenario. There is also no modification in
the soft sector. However, now one has secondary real radiation of the massive quark pair in the jet function. In this
scenario, the use of the OS subtraction for the virtual secondary massive quark loops in the jet function is not appropriate, 
since in that way one cannot smoothly interpolate to the massless limit, desirable in this scenario. Instead one has to use
the $\overline{\rm MS}$ prescription. Therefore the jet function has an analogous treatment as the hard matching coefficient:
it is evolved with $n_l + 1$ active flavors between $\mu_J$ and $\mu_m$, where a threshold correction has to be included,
and $n_l$-evolution is used between $\mu_m$ and $\mu_S$\,:
\begin{align}
\frac{1}{\sigma_0} \frac{\df\sigma}{\df\tau} \,=\, & \, Q\,H^{(n_l+1)}(Q,m,\mu_H)\,
 U^{(n_l+1)}_H(Q,\mu_H,\mu_m) {\mathcal{M}_H(Q,m,\mu_m)}\, U^{(n_l)}_H(Q,\mu_m,\mu_S)\!\!
\int\! \df s  \!\!\int\!\! \df s' \!\!\!\int \!\df s''\!\!\!\int\! \df s''' J^{(n_l+1)}(s''',m,\mu_J)\nonumber\\
&\times   U^{(n_l+1)}_J(s''-s''',\mu_m,\mu_J) \, \mathcal{M}_J(s'-s'',m,\mu_m) \, U^{(n_l)}_J(s-s',\mu_S,\mu_m) \,  S^{(n_l)}\Big(Q\,\tau-\frac{s}{Q},\mu_S\Big).
\label{eq:diffsigmaIII}
\end{align}
Here only the soft function and $U^{(n_l)}_J$ depend on $\alpha_s^{(n_l)}$, whereas the rest of elements 
depend on $\alpha_s^{(n_l + 1)}$. The jet function satisfies the massless limit
\begin{align}
\lim_{m\to 0} J^{(n_l+1)}(s,m,\mu) = J^{(n_l+1)}(s,\mu)\,,
\end{align}
as already indicated above, and the decoupling limit of the soft function is again trivial. The jet threshold
function $\mathcal{M}_J$ depends on the
large rapidity logarithm $\log(m^2/\mu_J^2)$, which again exponentiates, and its required perturbative expressions for an 
overall N$^3$LL resummation are known.
\subsection{Scenario IV: $\mathbf{Q>Q\,\lambda>Q\,\lambda^{\!2}>m}$}
Now $m$ is below any other renormalization scale, and therefore the hard and jet functions look the same as in Scenario III.
Furthermore their running proceeds with $n_l + 1$ flavors from their respective scales to $\mu_S$, since $\mu_m$ is not
crossed during the evolution. Therefore no threshold coefficients are necessary in this scenario. On the other hand now
we have real radiation effects in the soft function, and in complete analogy to the jet function in Scenario III we switch
to the $\overline{\rm MS}$ scheme, such the massless limit is manifest. The factorization theorem looks very similar to
Eq.~(\ref{eq:diffsigma0}), with the exception that the matrix elements depend explicitly on $m$:
\begin{align}
\frac{1}{\sigma_0}\frac{\df\sigma}{\df\tau}\,=\, &\, Q\,H^{(n_l+1)}(Q,m,\mu_H)\,
 \,U^{(n_l+1)}_H(Q,\mu_H,\mu_S) \int\! \df s\!  \int\! \df s' \,J^{(n_l+1)}(s',m,\mu_J)  \\
&\times U^{(n_l+1)}_J(s-s',\mu_S,\mu_J)\,S^{(n_l+1)}\Big(Q\,\tau-\frac{s}{Q},m,\mu_S\Big).\nn
\label{eq:diffsigmaIV}
\end{align}
Every single element of Eq.~(\ref{eq:diffsigmaIV}) depends on $\alpha_s^{(n_l + 1)}$ and all functions have a smooth massless
limit, in particular
\begin{align}
 \lim_{m\to 0}\hat{S}^{(n_l+1)}(\ell,m,\mu) = \hat{S}^{(n_l+1)}(s,\mu)\,,
\end{align}
therefore in this limit Eq.~(\ref{eq:diffsigmaIV}) exactly reduces to Eq.~(\ref{eq:diffsigma0}) with $n_f \to n_l + 1$.

Finally, in this scenario the evolution of the gap parameter \cite{Hoang:2008fs} also crosses the mass-mode threshold. 
Therefore its running 
proceeds in two steps and includes threshold corrections, in a similar way as for the strong coupling constant.
For the gap parameter one does not encounter any rapidity logarithms in the threshold corrections.
\section{Numerical Results}
\begin{figure}
\captionsetup{type=figure}
\subfloat[][]{
\includegraphics[width=0.49\columnwidth]{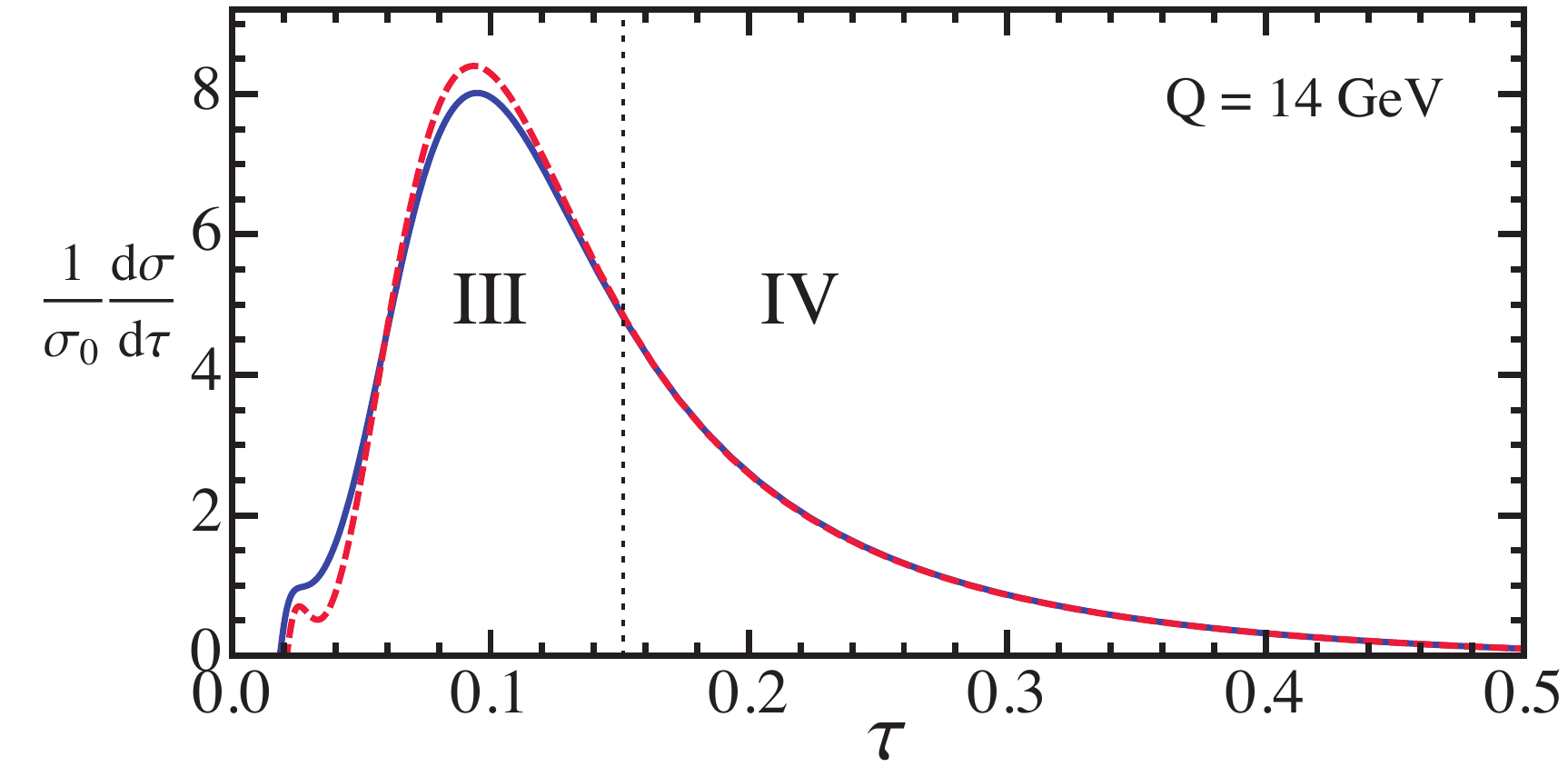}}~~~~
\subfloat[][]
{\includegraphics[width=0.48\columnwidth]{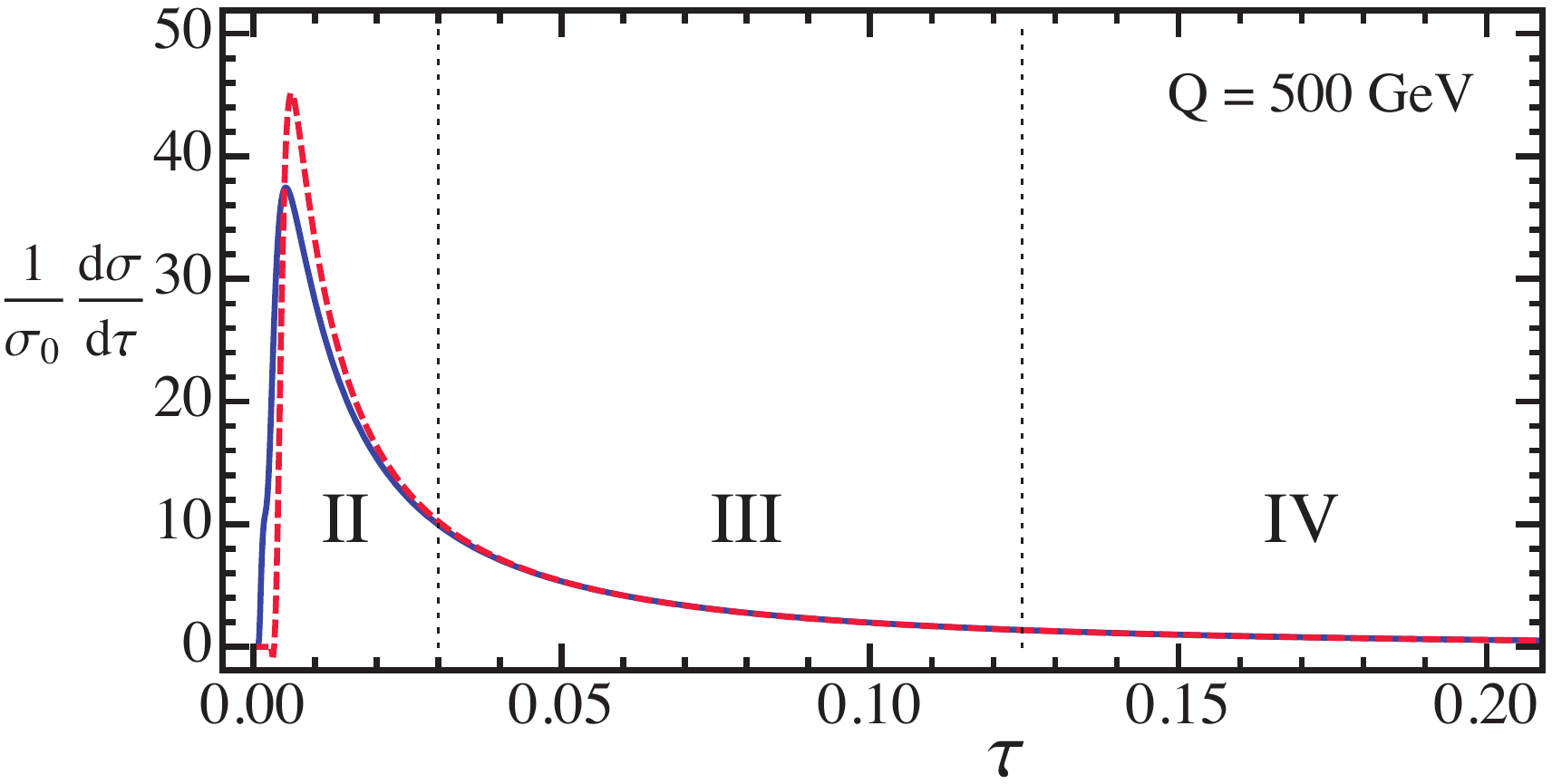}}
\caption{The thrust distribution for primary massless production at $Q=14$ GeV (a) and $Q = 500$ GeV (b) including secondary 
massive bottom (a) and top (b) effects (blue, solid) compared to keeping the bottom/top quark massless (red, dashed).
\label{fig:numerics}}
\end{figure}

In Fig.~\ref{fig:numerics} we show the effects of the secondary bottom (a) and top mass effects (b) on the thrust distribution,
at $14$~GeV and $500$~GeV, respectively. The plots only show the most singular terms of the distribution, and do not include
an estimate of the perturbative uncertainties. The renormalization scales have been chosen such that no large logs appear,
and their specific form can be found in Ref.~\cite{Pietrulewicz:2014qza}, as well as further plots. Hadronization effects
and renormalon subtractions are included. The different scenarios are indicated by horizontal dashed lines. One can see that
the effects are more visible in the peak of the distribution, and very small in the tail, where measurements of $\alpha_s$
are usually carried out.
\section{Conclusions}
In this work an explicit realization of a VFNS for final jet states has been presented, taking thrust as a case example.
It is suitable to include collinear and soft mass modes in the EFT setup to properly account for secondary heavy quark 
radiation, which allows to write down a set of factorization theorems. Depending on the relation between the mass scale and 
the other renormalization scales one of the four different factorization scenarios is applied. The key to achieve a continuous
description of the distribution for infinitely heavy masses (decoupling limit) to infinitesimally small ones (massless
limit) is to include the mass modes as passive or active running degrees of freedom related to the renormalization
subtractions corresponding to either the OS or $\overline{\rm MS}$ scheme, respectively, depending on the situation one is
facing. Real radiation effects in the jet and soft functions are included whenever they are kinematically allowed.
We have shown that the numerical impact of the secondary quark mass effects on the thrust distribution is small in the
tail region, but sizable at the peak. We plan to employ the VFNS setup for final state jets discussed here for a number of 
other applications where quark mass effects are more sizeable.

\begin{theacknowledgments}
We thank the Erwin-Schrdinger Institute (ESI) for partial support in the framework of the ESI program ``Jets and Quantum
Fields for LHC and Future Colliders''.
\end{theacknowledgments}

\bibliographystyle{aipproc}

\bibliography{../VFNSthrust}

\end{document}